\documentclass[10pt]{article}

\usepackage{amsmath,amsbsy,amssymb,amsthm}
\usepackage{a4wide}
\usepackage{graphicx}
\usepackage{longtable}
\usepackage{cite}

\begin{document}

\title{Pressure Responses of a Vertically Hydraulic Fractured Well\\
in a Reservoir with Fractal Structure\thanks{This is a preprint of a paper 
whose final and definite form will be published in 
\emph{Applied Mathematics and Computation}, ISSN: 0096-3003. 
Submitted 01/July/2014; revised 20/Dec/2014; accepted 29/Dec/2014.}}

\author{Kambiz Razminia$^{a}$\\
{\small \texttt{kambiz.razminia@gmail.com}}
\and
Abolhassan Razminia$^{b}$\\
{\small \texttt{razminia@pgu.ac.ir}}
\and Delfim F. M. Torres$^{c}$\\
{\small \texttt{delfim@ua.pt}}}

\date{$^{a}${\em{Department of Petroleum Engineering,\\
Petroleum University of Technology, Ahwaz, Iran}}\\[0.3cm]
$^{b}${\em{Dynamical Systems \& Control (DSC) Research Lab.,
Department of Electrical Engineering, School of Engineering,
Persian Gulf University,\\
P.O. Box 75169, Boushehr, Iran}}\\[0.3cm]
$^{c}${\em{\text{Center for Research and Development in Mathematics and Applications (CIDMA)},
Department of Mathematics, University of Aveiro, 3810--193 Aveiro, Portugal}}}

\maketitle

% ----------------------------------------------------------------

\begin{abstract}
We obtain an analytical solution for the pressure-transient behavior
of a vertically hydraulic fractured well in a heterogeneous reservoir.
The heterogeneity of the reservoir is modeled by using the concept of fractal geometry.
Such reservoirs are called fractal reservoirs. According to the theory
of fractional calculus, a temporal fractional derivative is applied to incorporate
the memory properties of the fractal reservoir. The effect of different parameters
on the computed wellbore pressure is fully investigated by various synthetic examples.

\bigskip

\end{abstract}

% -----------------------------------------

\section{Introduction}

Hydraulic fracturing plays an important role in improving the productivity of damaged wells
and wells producing. The vertical plane fracture is created by injecting fluid into the
formation and then filling with proponing agents, such as propants, to prevent closure.
In practical terms, two types of fractured well are considered:
infinite (high) or finite (low) conductivity vertical fracture. In case of infinite conductivity fracture,
it is assumed that the fluid flows along the fracture without any pressure drop.
Finite conductivity fracture occurs when the pressure drop along the fracture plane is not negligible.

The classical diffusion equation has been used to explain the pressure responses
of a well in a reservoir, which is assumed to be homogenous at all scales. However, recent
studies show that the homogeneity assumption is not valid in most cases \cite{R4,new:r1,R1,R5,R2,R3,new:r2}.
Due to this fact, fractal geometry has been used as an effective tool to describe the heterogeneities
of these reservoirs, which are called fractal reservoirs \cite{new:r3,R6,R7,R13}.
Since the diffusion process of fractal reservoirs is history dependent, and the anomalous
diffusion properties of fractal reservoirs cannot be fully described by the fractal model,
the concept of fractional derivative has been used to incorporate
the memory of the fluid flow \cite{R8,R9,R10,R11}.
Our main aim here is to analyze the pressure behavior of a well
with an infinite conductivity vertical fracture in a fractal reservoir.
An infinite radial system is considered in order to analyze the effects
of different parameters on the well response.

The paper is organized as follows. After this brief introduction,
the mathematical model is formulated in Section~\ref{sec2}.
A summary of the nomenclature used appears in Appendix~\ref{sec:append}.
The analytical solution to the model is provided in Section~\ref{sec3}.
Section~\ref{sec4} discusses how the well responses to different reservoir parameters.
The main conclusions of our study are given in Section~\ref{sec5}.

% ----------------------------------------

\section{Model description}
\label{sec2}

\begin{figure}[!ht]
\begin{center}
\includegraphics[scale=0.65]{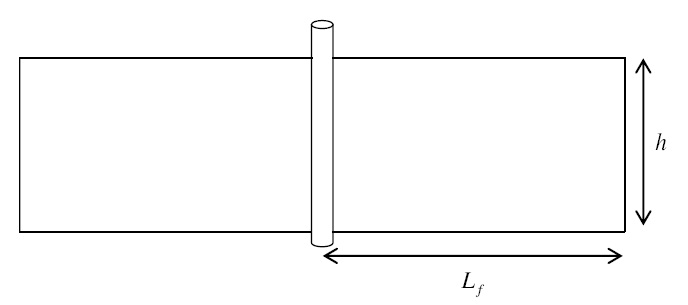}
\end{center}
\caption{Geometry of a vertically hydraulic fractured well.}
\label{Fig1}
\end{figure}

\begin{figure}[!ht]
\begin{center}
\includegraphics[scale=0.65]{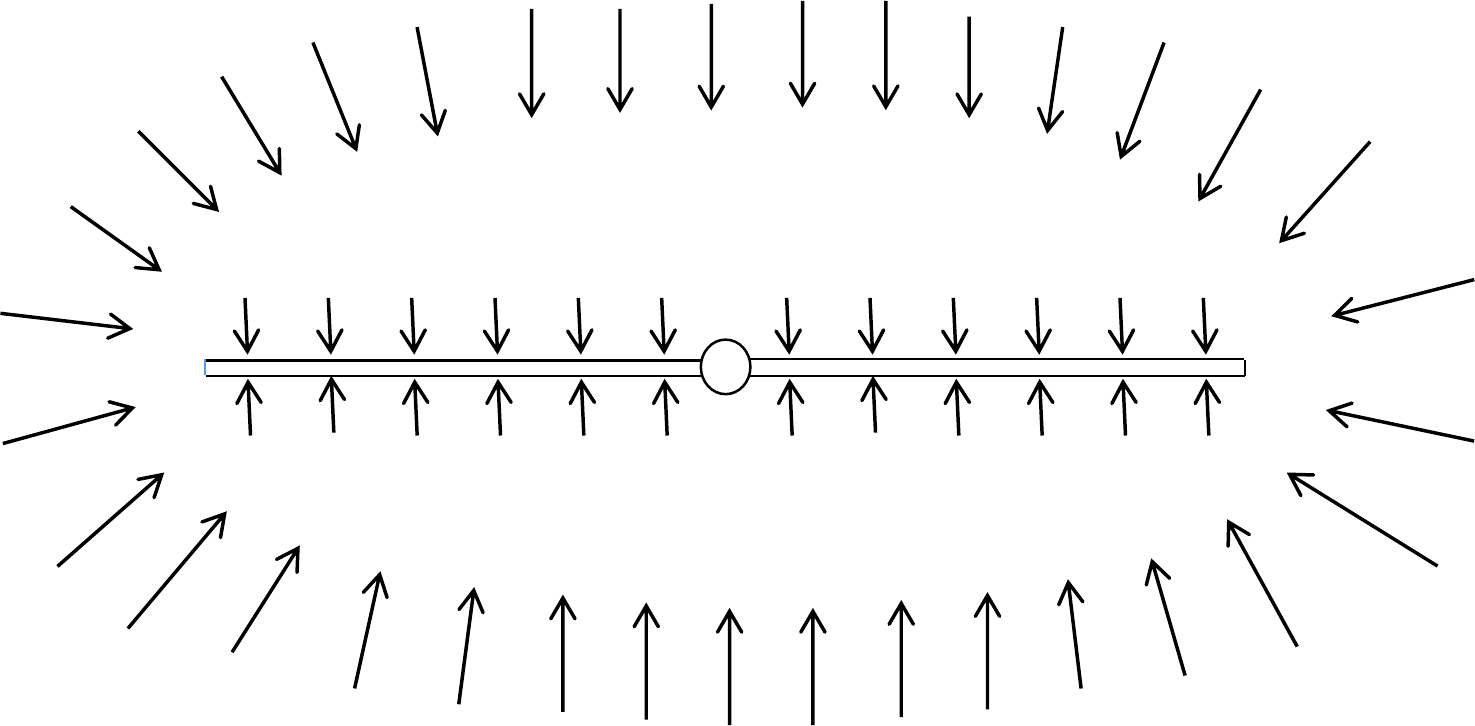}
\end{center}
\caption{Linear and pseudo-radial flow regimes near an infinite conductivity fracture.}
\label{Fig2}
\end{figure}

A schematic diagram of a vertically hydraulic fractured well is shown in Figure~\ref{Fig1}.
% ------------------------
% ------------------------
Figure~\ref{Fig2} shows the geometry of flow lines near the fractured well.
% ------------------------
% ------------------------
Before discussing the mathematical model of transport process, we define three variables:
\begin{itemize}
\item the dimensionless pressure
\[
{p_D} = \frac{{2\pi {\kern 1pt} k_w {\kern 1pt} h{\kern 1pt} ({p_i} - p(r,t))}}{{q\mu }};
\]

\item the dimensionless time
\[
{t_D} = \frac{{{k_w}{\kern 1pt} {r_{wD}}^\theta t}}{{{\phi _w}
{\kern 1pt} \mu {\kern 1pt} c{\kern 1pt} {{(2{L_f})}^2}}};
\]

\item the dimensionless radius
\[
{r_D} = \frac{r}{{2{L_f}}}.
\]
\end{itemize}
See Appendix~\ref{sec:append} for the description of all the quantities involved.
The best known and most useful model to describe the pressure behavior
of fractal reservoirs was firstly proposed by Metzler et al. \cite{R4}.
Similarly to Camacho-Vel\'{a}zquez et al. \cite{R5},
from here on we call \emph{generalized diffusion equation}
to the fractal-fractional diffusion (FFD) equation
\begin{equation}
\label{eq:1}
\frac{1}{{r_D^\theta }}\frac{{{\partial ^2}{p_D}}}{{\partial r_D^2}}
+ \frac{\beta }{{r_D^{\theta  + 1}}}\frac{{\partial {p_D}}}{{\partial {r_D}}}
= \frac{{\partial {}^\gamma {p_D}}}{{\partial {t_D}^\gamma}},
\end{equation}
where $\beta  = {d_{mf}} - \theta  - 1$. The parameter ${d_{mf}}$ denotes
the mass fractal dimension, while $\theta$ represents the conductivity index.
Mass fractal dimension is responsible for the reservoir structure,
and the conductivity index explains the diffusion process in the reservoir.
The Caputo fractional order derivative is used to introduce
${{{\partial ^\gamma }{p_D}} \mathord{\left/{\vphantom {{{\partial^\gamma }{p_D}}{\partial
{t_D}^\gamma }}} \right.\kern-\nulldelimiterspace} {\partial {t_D}^\gamma }}$:
\begin{equation}
\label{eq:2}
\frac{{{\partial ^\gamma }{p_D}}}{{\partial {t_D}^\gamma }}
= \frac{1}{{\Gamma (m - \gamma )}}\int\limits_0^{{t_D}}{{{({t_D} - \tau )}^{m
- \gamma  - 1}}{p_D}^{(m)}(\tau )d\tau},
\end{equation}
where $\gamma  \in {{\mathbb{R}}^+}$,
$\left\lceil \gamma  \right\rceil  = m \in {{\mathbb{Z}}^+}$,
and $\Gamma$ represents the Gamma function, that is,
\begin{equation}
\label{eq:25}
\Gamma(\nu) = \int_0^\infty  {{e^{ - \tau }}{\tau ^{\nu  - 1}}d\tau}.
\end{equation}

The order $\gamma$ of the fractional derivative
is related to the conductivity index by
$\gamma  = {2 \mathord{\left/{\vphantom {2 {(2 + \theta )}}} \right.
\kern-\nulldelimiterspace} {(2 + \theta )}}$.

% ----------------------------------------

\section{Analytical solution}
\label{sec3}

It is assumed that the pressure distribution of the reservoir
is uniform and constant at initial time:
\begin{equation}
\label{eq:3}
{p_D}({r_D},0) = 0.
\end{equation}
To obtain the line source solution of equation \eqref{eq:1},
we take ${r_w} \to {0^+}$. The inner boundary condition without
wellbore storage and skin effects can then be written as
\begin{equation}
\label{eq:4}
{r_D}^\beta {\left. {\frac{{\partial {p_D}({r_D},{t_D})}}{{\partial
{r_D}}}} \right|_{{r_D} \to {0^ + }}} =  -1.
\end{equation}
The outer boundary condition for the infinite reservoir is given by
\begin{equation}
\label{eq:5}
\mathop {\lim }\limits_{{r_D} \to \infty } \,{p_D}({r_D},{t_D}) = 0.
\end{equation}
Taking the Laplace transform to both sides of \eqref{eq:1},
and then using \eqref{eq:3}, we obtain that
\begin{equation}
\label{eq:6}
\frac{1}{{r_D^\theta }}\frac{{{\partial ^2}{{\bar p}_D}}}{{\partial r_D^2}}
+ \frac{\beta }{{r_D^{\theta  + 1}}}
\frac{{\partial {{\bar p}_D}}}{{\partial {r_D}}} = {s^\gamma }{\bar p_D},
\end{equation}
where $s$ is the Laplace transform variable. The dependent variable
${\bar{p}_D}$ denotes the Laplace transform of ${p_D}$,
and is a function of ${r_D}$ and $s$. In the Laplace space,
the inner boundary condition takes the form
\begin{equation}
\label{eq:7}
{r_D}^\beta {\left. {\frac{{\partial {{\bar p}_D}({r_D},s)}}{{\partial
{r_D}}}} \right|_{{r_D} \to {0^ + }}} =  - \frac{1}{s}
\end{equation}
while the outer boundary condition is given by
\begin{equation}
\label{eq:8}
\mathop {\lim }\limits_{{r_D} \to \infty } \,{\bar p_D}({r_D},s) = 0.
\end{equation}
Using the substitutions ${\bar p_D} = {r_D}^{(1 - \beta )/2}\bar W$
and $x={(2{s^{\gamma/2}}{r_D}^{(2 + \theta )/2})}/{(2 + \theta )}$,
and after a slight manipulation, we conclude that \eqref{eq:6} is equivalent to
\begin{equation}
\label{eq:9}
{x^2}\frac{{{\partial ^2}\bar W}}{{\partial {x^2}}}
+ x\frac{{\partial \bar W}}{{\partial x}} - ({x^2} + {\nu ^2})\bar W = 0,
\end{equation}
where $\nu  = \frac{{1 - \beta }}{{2 + \theta }}$.
Equation \eqref{eq:9} is Bessel's equation,
which has the general solution
\begin{equation}
\label{eq:10}
\bar W(x,s) = A{I_\nu }(x) + B{K_\nu }(x).
\end{equation}
Thus, the dimensionless pressure function in Laplace space can be written as
\begin{equation}
\label{eq:11}
{\bar p_D}({r_D},s) = {r_D}^{(1 - \beta )/2}\left[ {A{I_\nu }\left( {\frac{{2{s^{\gamma /2}}}}{{2
+ \theta }}{r_D}^{(2 + \theta )/2}} \right) + B{K_\nu }\left( {\frac{{2{s^{\gamma /2}}}}{{2
+ \theta }}{r_D}^{(2 + \theta )/2}} \right)} \right].
\end{equation}
Application of the outer boundary condition \eqref{eq:8} to \eqref{eq:11} yields $A = 0$.
Therefore, \eqref{eq:11} reduces to
\begin{equation}
\label{eq:12}
{\bar p_D}({r_D},s) = B{r_D}^{(1 - \beta )/2}{K_\nu }\left(
{\frac{{2{s^{\gamma /2}}}}{{2 + \theta }}{r_D}^{(2 + \theta )/2}}\right).
\end{equation}
Based on \eqref{eq:12}, the inner boundary condition \eqref{eq:7} can be written as
\begin{equation}
\label{eq:13}
- B\,\mathop {\lim }\limits_{{r_D} \to {0^ + }} {s^{\gamma /2}}{r_D}^{(1 + \beta
+ \theta )/2}{K_{\nu  - 1}}\left( {\frac{{2{s^{\gamma /2}}}}{{2
+ \theta }}{r_D}^{(2 + \theta )/2}} \right) =  - \frac{1}{s}.
\end{equation}
Equation \eqref{eq:13} can be simplified by using the formula
\begin{equation}
\label{eq:14}
{K_\nu }(x) \approx \frac{{\Gamma (\nu )}}{2}{\left( {\frac{2}{x}} \right)^\nu },
\quad \nu  > 0,
\end{equation}
valid for small arguments $0 < x \ll \sqrt {1 + \nu }$. Indeed, having in mind that
${K_\nu }(x) = {K_{ - \nu }}(x)$, and by making use of \eqref{eq:14},
we reduce \eqref{eq:13} to the following expression:
\begin{equation}
\label{eq:15}
B\,\mathop {\lim }\limits_{{r_D} \to {0^ + }} {s^{\gamma /2}}\frac{{\Gamma(1
- \nu )}}{{{2^\nu }}}{\left( {\frac{{2
+ \theta }}{{2{s^{\gamma/2}}}}}\right)^{1 - \nu }}
= \frac{1}{s}.
\end{equation}
Consequently,
\begin{equation}
\label{eq:16}
B\, = \frac{2}{{\Gamma (1 - \nu ){{(2 + \theta )}^{1 - \nu }}{s^{1 + \nu \gamma /2}}}}.
\end{equation}
Equation \eqref{eq:12} can be written as
\begin{multline}
\label{eq:17}
{\bar p_D}({x_D},{y_D},s) = B{\left( {\sqrt {{{({x_D} - {\alpha _1})}^2}
+ {{({y_D} - {\alpha _2})}^2}} } \right)^{(1 - \beta )/2}}\\
\times {K_\nu }\left({\frac{{2{s^{\gamma /2}}}}{{2
+ \theta }}{{\left( {\sqrt {{{({x_D} - {\alpha _1})}^2}
+ {{({y_D} - {\alpha _2})}^2}} } \right)}^{(2 + \theta )/2}}}\right).
\end{multline}
We obtain the pressure drop function, in a fractal reservoir
with a hydraulic fracture across the well,
by integrating \eqref{eq:12} with respect to ${\alpha_1}$
from $-1/2$  to $1/2$:
\begin{multline}
\label{eq:18}
{\bar p_{Df}}({x_D},{y_D},s) = B\int_{ - {1 \mathord{\left/
{\vphantom {1 2}} \right.
\kern-\nulldelimiterspace} 2}}^{{1 \mathord{\left/
{\vphantom {1 2}} \right.
\kern-\nulldelimiterspace} 2}} \Biggl[
{{{\left( {\sqrt {{{({x_D} - {\alpha _1})}^2}
+ {{({y_D} - {\alpha _2})}^2}} } \right)}^{(1 - \beta )/2}}}\\
\times {K_\nu }\left(
{\frac{{2{s^{\gamma /2}}}}{{2 + \theta }}{{\left( {\sqrt {{{({x_D} - {\alpha _1})}^2}
+ {{({y_D} - {\alpha _2})}^2}} } \right)}^{(2 + \theta )/2}}} \right)\Biggr] d{\alpha _1}.
\end{multline}
Since the computation of the pressure along the fracture is favorable,
it can be assumed that ${y_D} = {\alpha _2}$. Thus, the pressure drop function
\eqref{eq:18} reduces to
\begin{equation}
\label{eq:19}
{\bar p_D}({x_D},0,s) = B\int_{ - {1 \mathord{\left/
{\vphantom {1 2}} \right.
\kern-\nulldelimiterspace} 2}}^{{1 \mathord{\left/
{\vphantom {1 2}} \right.
\kern-\nulldelimiterspace} 2}} {{{\left( {\sqrt {{{({x_D}
- {\alpha _1})}^2}} } \right)}^{(1 - \beta )/2}}{K_\nu }\left(
{\frac{{2{s^{\gamma /2}}}}{{2 + \theta }}{{\left( {\sqrt {{{({x_D}
- {\alpha _1})}^2}} } \right)}^{(2 + \theta )/2}}} \right)d{\alpha _1}},
\end{equation}
where $\left| {{x_D}} \right|
\le {1 \mathord{\left/{\vphantom {1 2}}
\right.\kern-\nulldelimiterspace} 2}$.
By changing variables, and after further manipulations,
\eqref{eq:19} can be written as
\begin{multline}
\label{eq:20}
{\bar p_D}({x_D},0,s) = B\frac{1}{{{s^{\gamma /2}}}}{\left( {\frac{{2
+ \theta }}{{2{s^{\gamma /2}}}}} \right)^{a - 1}}
\Biggl(
\int_0^{\left( {{{2{s^{\gamma /2}}} \mathord{\left/
{\vphantom {{2{s^{\gamma /2}}} {(2 + \theta )}}} \right.
\kern-\nulldelimiterspace} {(2 + \theta )}}} \right){{\left( {{1 \mathord{\left/
{\vphantom {1 2}} \right.
\kern-\nulldelimiterspace} 2} - {x_D}} \right)}^{{{(2 + \theta )} \mathord{\left/
{\vphantom {{(2 + \theta )} 2}} \right.
\kern-\nulldelimiterspace} 2}}}} {{z^{a - 1}}{K_\nu }\left( z \right)dz}\\
+ \int_0^{\left( {{{2{s^{\gamma /2}}} \mathord{\left/
{\vphantom {{2{s^{\gamma /2}}} {(2 + \theta )}}} \right.
\kern-\nulldelimiterspace} {(2 + \theta )}}} \right){{\left( {{1 \mathord{\left/
{\vphantom {1 2}} \right.
\kern-\nulldelimiterspace} 2} + {x_D}} \right)}^{{{(2 + \theta )} \mathord{\left/
{\vphantom {{(2 + \theta )} 2}} \right.
\kern-\nulldelimiterspace} 2}}}} {{z^{a - 1}}{K_\nu }\left( z \right)dz} \Biggr)
\end{multline}
with $a = {{(4 - {d_{mf}} + \theta )} \mathord{\left/{\vphantom{{(4
- {d_{mf}} + \theta )} {(2 + \theta )}}} \right.\kern-\nulldelimiterspace} {(2 + \theta )}}$.
The integrals in \eqref{eq:20} can be computed by the following formula:
\begin{multline}
\label{eq:21}
\int {{z^{a - 1}}{K_\nu }(z)dz}
=  - \frac{{{2^{\nu  - 1}}\pi {z^{a - \nu }}\csc (\pi \nu )}}{{(\nu  - a)
\Gamma (1 - \nu )}}{}_1{F_2}\left( {\frac{{a - \nu }}{2};1 - \nu ,
\frac{{a - \nu }}{2} + 1;\frac{{{z^2}}}{4}} \right)\\
- \frac{{{2^{ - \nu  - 1}}\pi {z^{a + \nu }}\csc (\pi \nu )}}{{(a + \nu )
\Gamma (1 + \nu )}}{}_1{F_2}\left( {\frac{{a + \nu }}{2};1
+ \nu ,\frac{{a + \nu }}{2} + 1;\frac{{{z^2}}}{4}} \right),
\end{multline}
where $\nu  \notin \mathbb{Z}$ and ${_1F_2}$ represents
the generalized hypergeometric function. For the Euclidean model,
that is, the particular case $\nu  = 0$  and $a = 1$,
the presented integrals can be evaluated by
\begin{equation}
\label{eq:22}
\int {{K_0}(z)dz}  = \frac{{\pi z}}{2}\left(
{{K_0}(z){L_{ - 1}}(z) + {K_1}(z){L_0}(z)} \right)
\end{equation}
or
\begin{equation}
\label{eq:23}
\int {{K_0}(z)dz}  = \frac{{\pi z}}{2}\left(
{{K_0}(z){L_1}(z) + {K_1}(z){L_0}(z)} \right) + z{K_0}(z),
\end{equation}
where ${L_{ - 1}}(z)$, ${L_0}(z)$ and ${L_1}(z)$
are modified Struve functions given by
\begin{equation}
\label{eq:24}
{L_\nu }(z) = \frac{{{z^{\nu  + 1}}}}{{{2^\nu }\sqrt \pi
\Gamma \left( {\nu  + \frac{3}{2}} \right)}}\,
{_1F}_2\left( {1;\frac{3}{2},\nu
+ \frac{3}{2};\frac{{{z^2}}}{4}} \right),
\quad - \nu - \frac{3}{2} \notin \mathbb{N}.
\end{equation}
The inverse transform operation is easily carried out
by using the Gaver--Stehfest algorithm \cite{R12}.

% ----------------------------------------

\section{Model responses}
\label{sec4}

\begin{figure}[!ht]
\begin{center}
\includegraphics[scale=0.45]{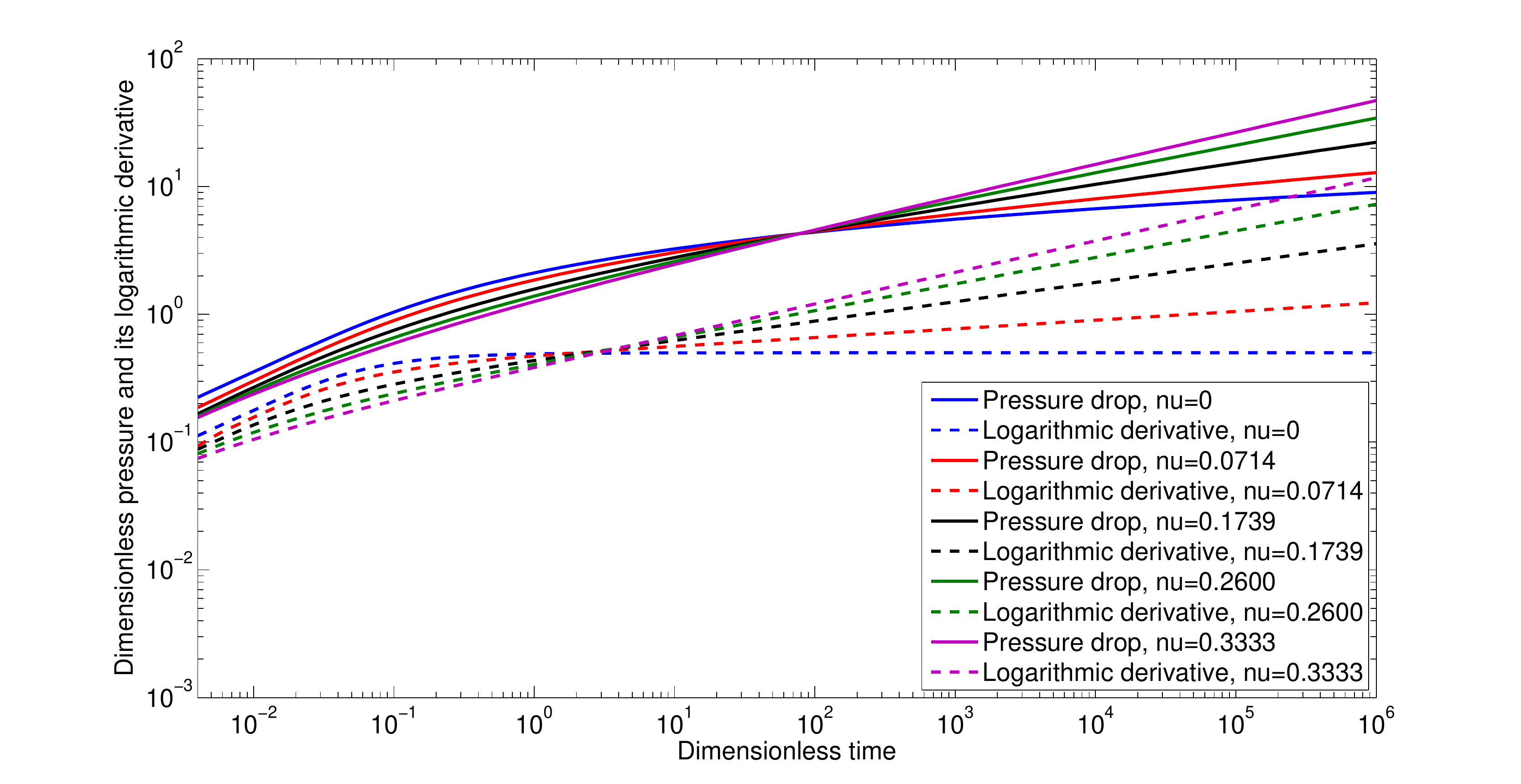}
\end{center}
\caption{Responses for a fractured well without wellbore
storage and skin effects. Log-log analysis of $p_D$ and
${d{p_D}}/{d\ln ({t_D})}$ versus $t_D$ with
${d_{mf}} = \{ 2,\,1.95,\,1.9,\,1.85,\,1.8\}$,
$\theta  = \{ 0,\,0.1,\,0.3,\,0.5,\,0.7\}$
and $\gamma  = \{ 1,\,0.9524,\,0.8696,\,0.8000,\,0.7407\}$.}
\label{Fig3}
\end{figure}

\begin{figure}[!ht]
\begin{center}
\includegraphics[scale=0.45]{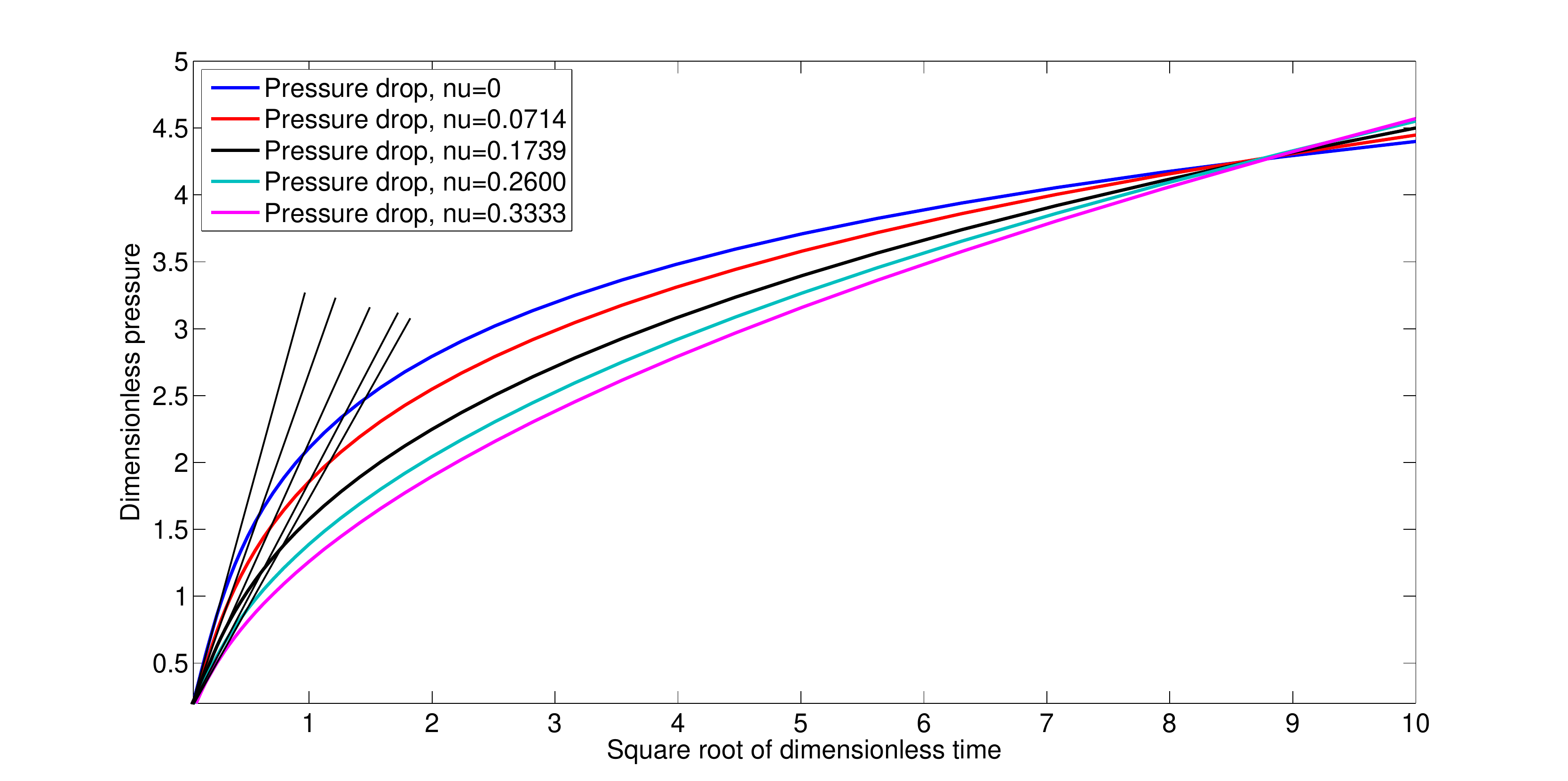}
\end{center}
\caption{Specialized analysis of $p_D$ versus $\sqrt{{t_D}}$
with ${d_{mf}} = \{ 2,\,1.95,\,1.9,\,1.85,\,1.8\}$,
$\theta  = \{ 0,\,0.1,\,0.3,\,0.5,\,0.7\}$ and
$\gamma  = \{ 1,\,0.9524,\,0.8696,\,0.8000,\,0.7407\}$.}
\label{Fig4}
\end{figure}

To show the pressure-transient behavior of a fractured well,
we use \eqref{eq:20} to compute the wellbore pressure response
at ${x_D} = 0$ (uniform flux solution), without wellbore storage and skin effects.
The well intercepts the center of a single vertical fracture plane.
According to the uniform flux fracture assumption, the flow rate per unit
of fracture surface is constant along the fracture length.
Figure~\ref{Fig3} shows the different behavior of a well
response in a homogenous and fractal reservoir.
Generally, the fracture structure works as a sink,
which enforces the fluid to go towards the fracture
with hyperbolical flow geometry. The geometry of fracture
enforces the flow lines to be perpendicular to the fracture plane
and the pressure-transient response defines a linear flow in the reservoir.
After the linear flow regime, when the effect of fluid flow reaches
the two ends of the fracture, the geometry of flow can be identified,
for a small time, by a hyperbolical flow regime. Ultimately,
the characteristic radial flow regime is observed from the well response.
The linear flow regime can be identified by two straight-lines of pressure
and its logarithmic derivative at early times (see Figure~\ref{Fig3}). As can be seen
from Figure~\ref{Fig3}, these two parallel straight-lines have a half
slope during the linear flow regime in a conventional reservoir.
The specialized analysis (Cartesian coordinates) on Figure~\ref{Fig4}
shows that with a plot of the pressure change versus the square root of the elapsed time,
the linear flow can be identified with a straight-line intercepting the origin.
The results of Figure~\ref{Fig4} show that the slope of the straight-line
is decreasing with the increasing of the parameter $\nu$.
In other words, in a more complex reservoir,
described by a larger value of the conductivity index $\theta$
and a smaller value of the mass fractal dimension ${d_{mf}}$, the linear flow
is identified with a smaller slope of the straight-line in the specialized plot.
Figure~\ref{Fig3} indicates that the radial flow regime (at late times)
is identified by the straight-line shape of the logarithmic derivative in log-log scale
and that the reservoirs with more complex structures (larger values of $\nu$) have the straight-lines
with larger slopes. Moreover, Figure~\ref{Fig3} shows that in a fractal reservoir,
at late times during the radial flow regime, the pressure response and its
logarithmic derivative are parallel straight-lines. This fact indicates that
the pressure and its logarithmic derivative can be expressed by
two power-law functions with the same power.
% ------------------------
% ------------------------
% ------------------------

\begin{figure}[!ht]
\begin{center}
\includegraphics[scale=0.45]{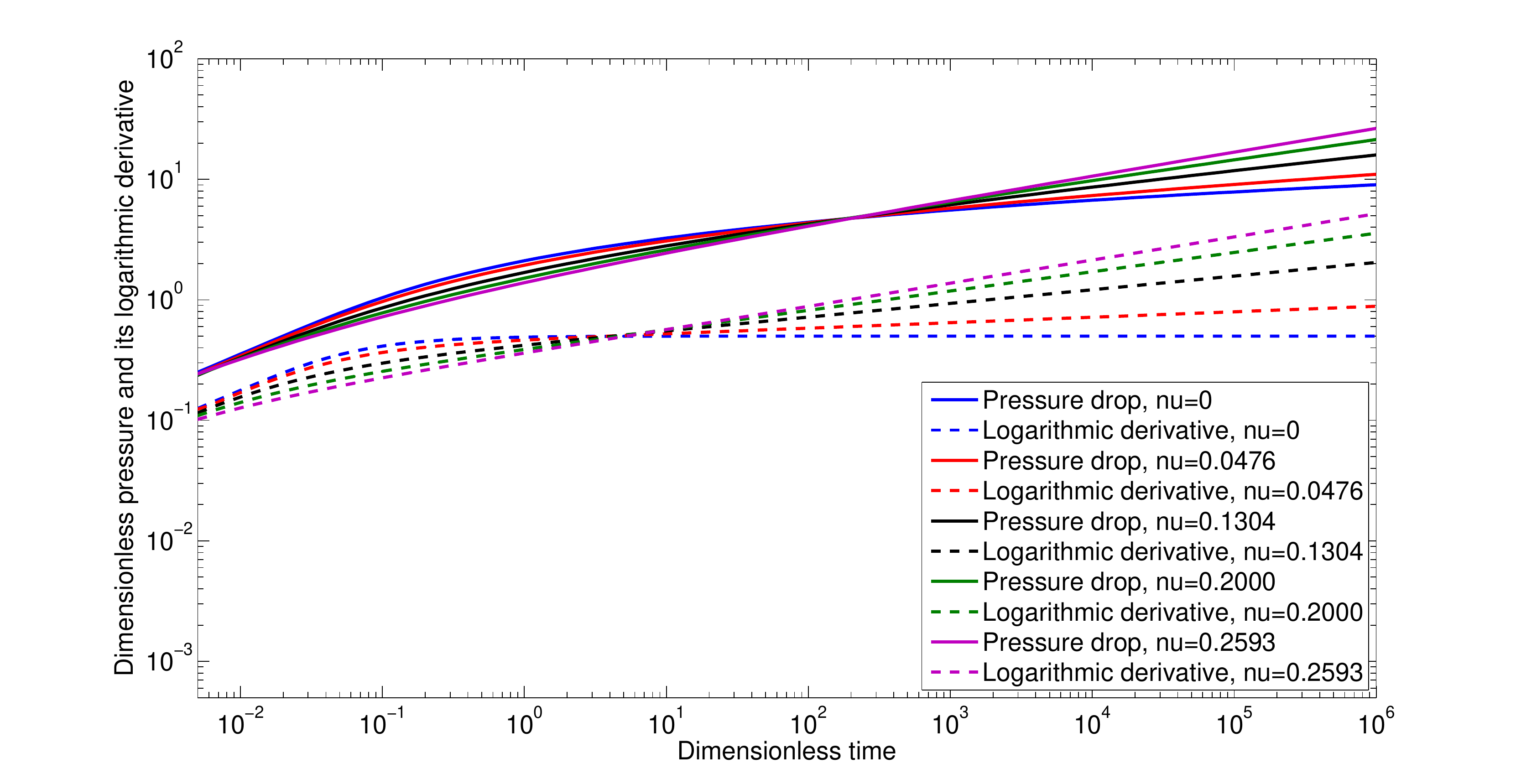}
\end{center}
\caption{Responses for a fractured well without wellbore storage and skin effects.
Log-log analysis of $p_D$ and $d{p_D}/{d\ln({t_D})}$ versus $t_D$ with
${d_{mf}} = 2$, $\theta  = \{ 0,\,0.1,\,0.3,\,0.5,\,0.7\}$
and $\gamma  = \{ 1,\,0.9524,\,0.8696,\,0.8000,\,0.7407\}$.}
\label{Fig5}
\end{figure}

The reservoir rock porosity is a statistical property of the system,
and can be expressed as
$\phi (r) = {\phi _r}{\left( {{r \mathord{\left/
{\vphantom {r {{L_r}}}} \right.
\kern-\nulldelimiterspace} {{L_r}}}} \right)^{{d_{mf}} - 2}}$,
where $\phi_r$ is a reference porosity value at the reference length $L_r$.
Particularly, the reference length is the wellbore radius, and the porosity
of a fractal reservoir can be written as
$\phi (r) = {\phi _w}{\left( {{r \mathord{\left/
{\vphantom {r {{r_w}}}} \right.
\kern-\nulldelimiterspace} {{r_w}}}} \right)^{{d_{mf}} - 2}}$.
The reference porosity value can be assumed to be equal to the porosities obtained
from log and core data. However, based on log and core porosities, the reasonable
oil-in-place estimates indicate that the porosity variations are not big,
and can be considered equal to the constant reference porosity.
It can be therefore immediately concluded that the mass fractal dimension ${d_{mf}}$
is equal to the Euclidean dimension, that is, ${d_{mf}} = 2$. On the other hand, because
$k(r) = {k_r}{\left( {{r \mathord{\left/
{\vphantom {r {{L_r}}}} \right.
\kern-\nulldelimiterspace} {{L_r}}}} \right)^{{d_{mf}} - \theta  - 2}}$ and,
particularly, $k(r) = {k_w}{\left( {{r \mathord{\left/
{\vphantom {r {{r_w}}}} \right.
\kern-\nulldelimiterspace} {{r_w}}}} \right)^{{d_{mf}} - \theta  - 2}}$,
the deficiencies of average permeability calculations indicate
that the dynamical properties of the system have an important role in the flowing
fluid through the reservoir and in all stages of production.
Since ${d_{mf}} = 2$, the permeability variations can be expressed by
$k(r) = {k_w}{\left( {{r \mathord{\left/{\vphantom {r {{r_w}}}} \right.
\kern-\nulldelimiterspace} {{r_w}}}} \right)^{-\theta}}$. So, it requires investigating
the effect of the conductivity index $\theta$ variations on the well response.
As can be seen from Figure~\ref{Fig5}, by increasing the $\theta$ values, the pressure
responses and their logarithmic derivatives increase. For a constant-rate production,
more pressure drops in reservoirs with a more complicated diffusion process
(larger $\theta$) indicates that the diffusion is slower in these types of reservoirs.
Therefore, application of the conventional model solutions to such reservoirs
may lead the analyst to a wrong interpretation.
% ------------------------
% ----------------------------------------

\section{Concluding remarks}
\label{sec5}

We used fractional calculus to model an infinite conductivity vertically fractured well
in a fractal reservoir with more realistic behavior than the conventional (homogeneous) reservoirs.
An analytical solution was obtained, allowing us to analyze and interpret
the well test data taken from these types of wells and reservoirs.
On the basis of our results, the simulated examples and the discussion presented,
the following conclusions can be expressed:
\begin{itemize}
\item the Euclidean interpretation model cannot be used to analyze
the pressure-transient behavior of fractal reservoirs;

\item the obtained solution can be used as a practical tool to analyze
the effects of dynamical properties of the system;

\item according to the shortcomings of the average permeability estimates,
the analytical solution can be used to determine
the conductivity index parameter $\theta$
and then the permeability variations.
\end{itemize}

% ----------------------------------------

\appendix

% ----------------------------------------

\section{Nomenclature}
\label{sec:append}

\begin{center}
\begin{longtable}{|l|l|l|}
\hline \multicolumn{1}{|c|}{\textbf{Symbol}} & \multicolumn{1}{c|}{\textbf{Meaning}}
& \multicolumn{1}{c|}{\textbf{Units}} \\ \hline
\endfirsthead

\multicolumn{3}{c}{{\bfseries \tablename\ \thetable{} -- continued from previous page}} \\
\hline \multicolumn{1}{|c|}{\textbf{Symbol}} & \multicolumn{1}{c|}{\textbf{Meaning}}
& \multicolumn{1}{c|}{\textbf{Units}} \\ \hline
\endhead

\hline \multicolumn{3}{|r|}{{Continued on next page}} \\ \hline
\endfoot

\hline
\endlastfoot

$c$ & compressibility & vol/vol/atm\\
${d_{mf}}$ & mass fractal dimension &  \\
${}_1{F_2}$ & generalized hypergeometric function & \\
$h$ & reservoir thickness & cm\\
${I_\nu}$ & modified Bessel function of the first kind of order $\nu$ &  \\
$k$ & permeability & darcy\\
$k_w$ & permeability at the wellbore & darcy\\
$K_\nu$ & modified Bessel function of the second kind of order $\nu$ &  \\
$L_f$ & fracture half-length & cm\\
$L_\nu$ & modified Struve function of order $\nu$ & \\
$p$ & pressure & atm\\
$p_D$ & dimensionless pressure &  \\
$p_i$ & initial pressure & atm \\
$q$ & production rate & cc/sec\\
$r$ & radial distance & cm\\
$r_D$ & dimensionless radius &  \\
$r_w$ & wellbore radius & cm \\
$r_{wD}$ & dimensionless wellbore radius & \\
$t$ & time & sec\\
$t_D$ & dimensionless time &  \\
$s$ & Laplace image space variable &  \\
$\gamma$ & fractional derivative order &  \\
$\Gamma$ & Gamma function &  \\
$\theta$ & conductivity index &  \\
$\mu$ & viscosity & cP\\
$\phi$ & porosity (or void fraction) & \\
$\phi_w$ & porosity at the wellbore &
\end{longtable}
\end{center}

% ----------------------------------------

\section*{Acknowledgments}

This work was partially supported by CIDMA \& FCT
within project PEst-OE/MAT/UI4106/2014.

% ----------------------------------------

% ----------------------------------------

\end{document}